\documentclass[10pt,conference]{IEEEtran}
\IEEEoverridecommandlockouts
\usepackage{cite}
\usepackage{mathtools,xparse}
\usepackage{amsmath,amssymb,amsfonts}

\usepackage{geometry}

\usepackage{amsthm}
\usepackage{float}

\usepackage{cite}
\usepackage{relsize}
\usepackage{float}
\usepackage{mathtools}
\usepackage{graphics}
\usepackage{bm}
\usepackage{algorithmic}
\usepackage{graphicx}
\usepackage{textcomp}
\usepackage{xcolor}
\usepackage[pdfa]{}

\usepackage{mwe}

\newtheorem{Theo.}{Theorem}

\newtheorem{Lemma}{Lemma}

\def\Var{{\textrm{Var}}\,}

\ifCLASSINFOpdf
\else
 \DeclareGraphicsExtensions{.eps}
\fi
\begin{document}
	\onecolumn
	This work has been submitted to the IEEE for possible publication. Copyright may be transferred without notice, after which this version may no longer be accessible.
	%
\twocolumn
	\title{Performance Analysis and Optimization of 3D Massive MIMO Multi-Pair Relaying with SWIPT}

	\author{\IEEEauthorblockN{Samira Rahimian, Yindi Jing, Masoud Ardakani\\ }
		\IEEEauthorblockA{Department of Electrical and Computer Engineering, University of Alberta, Canada\\
			Email: {\{srahimia, yindi, ardakani\}}@ualberta.ca}
	}
	
	
	%


	\maketitle
\begin{abstract}
In this paper, we study a massive multiple-input multiple-output
(mMIMO) relay network where multiple source-destination pairs exchange information through a common relay equipped with a massive antenna array. The source users perform simultaneous wireless information and power transfer (SWIPT) and the power-splitting (PS) scheme is used at the relay to first harvest energy from the received signals, and then, to transmit the decoded signals using the harvested energy. Relay performs maximum ratio combining/maximum ratio transmission (MRC/MRT) beam-forming on the received signal. Under the three-dimensional (3D) directional channel model, we derive a closed-form lower bound expression for the average signal-to-interference-plus-noise ratio using results from random matrix theory, which leads to an asymptotic approximation of the achievable sum-rate. Based on that, we study a joint optimization problem over the tilt and PS ratio to maximize the achievable sum-rate. Grid search algorithm is used to solve the non-convex problem. Simulation results verify our theoretical analysis and the efficiency of our optimized design. In particular, our optimized system outperforms a conventional system with $\pi/4$ tilt and PS of $0.5$, by at least $61\%$.  

	\end{abstract}


	%
	\IEEEpeerreviewmaketitle

	\section{Introduction}

Massive multiple-input multiple-output (mMIMO) relaying has attracted lots of attention in 5G and beyond wireless systems as it provides notable spectral and energy efficiency, coverage and reliability \cite{andrews2014will,larsson2014massive}.  Also, with the advent of Internet-of-things (IoT) \cite{gelenbe2015impact}, an emerging solution for prolonging the lifetime of
energy constrained relays is simultaneous wireless information and power transfer (SWIPT). Through SWIPT, the received signal  can be split into two distinct parts: the information decoding (ID) and the energy harvesting (EH) \cite{krikidis2014simultaneous}. As a
practical technique for co-located receivers power splitting (PS), splits the received signal into two power levels for ID and EH\cite{zhang2013mimo}.  

For SWIPT in  multiple-input single-output systems, \cite{shi2015energy} has performed the optimization over the PS ratio assuming zero-forcing (ZF) beam-forming at the base station (BS). SWIPT for an amplify-and-forward (AF) one-way relay system with a single source-destination pair is considered in \cite{nasir2013relaying}, where two SWIPT protocols based on PS and time switching (TS) are proposed. Further, for the same setup, but with a decode-and-forward (DF) scheme, \cite{ju2015maximum} optimizes PS and TS ratios to maximize the transmission rate. In \cite{ding2014power}, for a multi-pair one-way relay network, two SWIPT based strategies to distribute the harvested energy among the users are investigated. 

For mMIMO relays, \cite{liu2016multipair} assumes
a DF multi-pair relay network, where the relay harvests energy from both the source and destination users based on the PS protocol. Maximum ratio combining/maximum ratio transmission (MRC/MRT) beam-forming is employed at the relay. Through asymptotic analysis, it is shown that the harvested energy is independent of the fast fading effect, and that the transmission power of each source and destination can be scaled inversely proportional to the number of relay antennas. In \cite{amarasuriya2016wireless}, asymptotic sum-rate analysis for a multi-way relay network where the relay harvests energy and employs ZF is provided. In \cite{wang2017wireless}, a two-way multi-pair relay network is considered where the users harvest energy. The users' PS ratios are optimized to maximize  the achievable rates assuming ZF and MRC. In \cite{wang2019performance}, an AF two-way relay network with a single pair of users where the relay harvests energy based on PS is assumed. The asymptotic sum-rate is analyzed and the optimum PS ratio is obtained considering ZF.  
 
Three-dimensional (3D) MIMO via employing active antenna systems in both horizontal
and vertical domains, is an appealing technology to improve the efficiency of both
information and energy transfer while dealing with physical constraints in mMIMO systems\cite{nam2013full,cheng2014communicating}. In SWIPT, the transmission-distance-to-BS-height ratio is
usually smaller than that in a conventional macro cell. Thus, the
vertical domain is as important as the horizontal domain. The
vertical domain has been utilized for SWIPT with 3D
sectorized antennas in \cite{krikidis2016swipt} where the performance is analyzed. Antenna tilt is a parameter for	adapting the elevation angle of the antenna pattern which has the potential to bring significant performance
gains. In \cite{fan2017exploiting}, SWIPT in a 3D mMIMO downlink is considered where the BS tilt, along with the users' PS ratios are optimized to minimize the BS transmit power. 

In this paper, we study a SWIPT enabled multi-user one-way mMIMO relay network with 3D directional  antennas. There has been no existing work on the performance analysis and optimization of such systems. Under the PS protocol and MRC/MRT at the relay, we derive a closed-form expression that serves as a lower-bound on the average signal-to-interference-plus-noise ratio (SINR),  using results for Haar matrices. Base on that, an asymptotic average achievable sum-rate expression is obtained. A joint optimization problem over the relay PS ratio and antenna array tilt is formulated to maximize the average achievable sum-rate. Grid search algorithm is used to solve the non-convex problem. Monte-Carlo simulation results are presented to verify our theoretical analysis and the gains that the optimized set-up brings. It is shown that the optimized system outperforms a conventional system with $\pi/4$ tilt and PS of $0.5$, by at least $62\%$.

 

	\section{System Model}\label{sec:sys}
The considered 3D multi-pair one-way relay network is shown in Fig. \ref{3D}. There are $K$ pairs of single-antenna users that are separated into two groups: source users $u_{Sk}$, and destination users $u_{Dk}$, for  $k=1,..., K$, communicating through a common mMIMO relay. $u_{Sk}$ sends information to $u_{Dk}$. The number of antennas at the relay is denoted by $N$. In the multiple access phase, the source users perform SWIPT and the relay uses the PS scheme, meaning that part of the received signal is used to decode the source information, while the other part is used to harvest energy for the coming broadcast phase. In the broadcast phase, the relay uses the harvested energy and transmits the decoded information to the destination users. The relay employs MRC/MRT beam-forming in the DF process.
\subsection{Relay 3D Antenna Pattern } \label{sec:sys_antenna}
\begin{figure}[!t]
	\vspace{-1cm}
	\begin{picture}(80,160)
	\put(-5,-37){\includegraphics[width=3.2in]{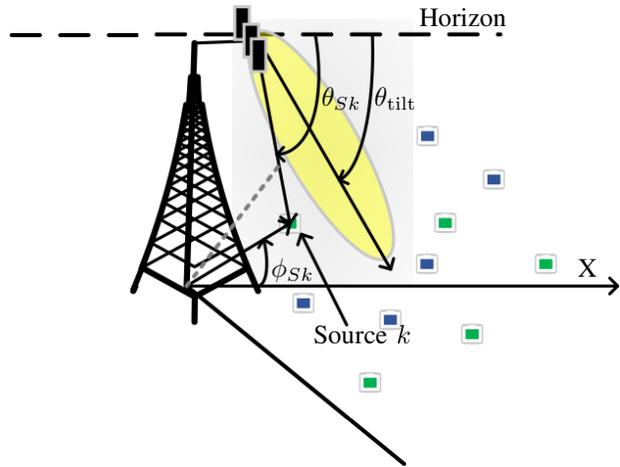}}
	\put(94,35){$\phi_{Sk}$}
	\put(113,100){$\theta_{Sk}$}
	\put(133,100){$\theta_{\mathrm{tilt}}$}
	\put(150,130){Horizon}
	\put(210,35){X}
	\put(110,10){Source $k$}
	\end{picture}
	\vspace{1cm}
	\caption{ The green squares represent the source users and the blue ones are the destination users. The spherical angles of the $k$th source user are illustrated. The coverage area in the horizontal plane spans an angular range of $120^{\circ}$.}
	\label{3D}
	\vspace{-.5cm}
\end{figure}

We assume that the relay antenna array is placed in the plane parallel to the ground and each antenna transmits the same signal with a specific weight.  By tuning the weights, the tilt can be controlled. We assume that a common tilt is applied at all antennas and approximate the antenna pattern using the 3D directional model  in 3GPP\cite{access2010further}. The observed antenna
 gain from any antenna of the relay at the $k$th source or destination user, for $i=S,D$, is expressed in dBi scale as follows:
\begin{align}
A_{ik}^{\text{dBi}}(\theta_{\text{tilt}})=&-\Big(\text{min}\Big[12\Big(\frac{\phi_{ik}}{\phi_{\text{3dB}}}\Big)^{2}, \text{SLL}_{\text{az}}\Big]+ \notag\\
&\text{min}\Big[12\Big(\frac{\theta_{ik}-\theta_{\text{tilt}}}{\theta_{\text{3dB}}}\Big)^{2}, \text{SLL}_{\text{el}}\Big]\Big), \label{eq:antenna_gain}
\end{align}
where  $0<\theta_{\rm{tilt}}<\pi/2$  is the tilt between the horizon and the beam peak, $\theta_{ik}$ is the angle between the
horizon and the line connecting the user to the relay antenna array, $\phi_{ik}$ is  the angle between the X-axis and the line in the horizontal plane connecting User $k$ to the projection point of the relay on the horizontal plane. A schematic illustration of these angles are shown in Fig.~\ref{3D}.   Moreover, $\text{SLL}_{\text{az}}=25$ dB and 
$\text{SLL}_{\text{el}}=20$ dB are the sidelobe levels (SLLs) of the antenna patterns in the
horizontal and vertical planes, respectively.  The  3-dB beamwidth in the horizontal and vertical planes are denoted as $\phi_{\text{3dB}}=65^{\circ}$ and $\theta_{\text{3dB}}=6^{\circ}$, respectively.  Further, it is assumed that the relay beam peak is fixed on $\phi=0$ relative to the X-axis. 
\subsection{ Channel Model}
We assume that a direct link between each pair of source and destination does not exist due to the path-loss. Let $\mathbf{G}_i=\mathbf{H}_i\mathbf{D}_i^{\frac{1}{2}}$ be the channel matrix from the users to the relay for $i\in\{S, D\}$, where $\mathbf{H}_i\in \mathbb {C}^{N\times K}$ is the small scale Rayleigh fading matrix whose elements are independent and identically distributed (i.i.d.), each following $\mathcal{CN}(0,1)$.  Further, the diagonal $K\times K$ matrix $\mathbf{D}_i=\mathrm{diag}\{\beta_{ik}(\theta_{\text{tilt}})\} $ for $ k \in \{1,\cdots K\}$, $i \in \{S,D\},$ accounts for the large scale fading coefficients, including path loss and antenna gain  given by,
\begin{equation}
\beta_{ik}(\theta_{\text{tilt}})=d_{ik}^{-\nu}A_{ik}(\theta_{\text{tilt}}), \label{eq:beta}
\end{equation}
where $d_{ik}$ denotes the distance between User $k$ and the relay, and $\nu$ is the path loss exponent. The channel fading keeps invariant in each relaying time block, but changes independently from one block to another. It is assumed that the relay has perfect CSI\footnote{The analytical method and result can be straightforwardly generalized to systems with CSI error.}. 
\vspace{-0.15cm}
\section{Average Achievable Sum-rate Analysis and Maximization}\label{sec:achiev}
All nodes work in the half-duplex mode. Transmissions are conducted with a two-phase protocol as explained in the following subsections.
\subsection{Phase I: Multiple Access (MAC) and SWIPT}
In phase I, the source users perform SWIPT and transmit their vector of information signals, $\mathbf{x},$ to the relay. The information signals are normalized for unit power as $\mathbb{E}\{|x_k|^2\}=1,  k\in\{1,2,\cdots, K\}$, where $x_k$ is the $k$th element of $\mathbf{x}$. The average transmit power of each source user is denoted as  $p_s$. Thus, the received signal at the relay is
\begin{align}
\mathbf{y}_{\mathrm{R}}=\sqrt{p_s}\mathbf{G}_S\mathbf{x}+\mathbf{n}^{\prime}_\mathrm{R},
\end{align}
where $\mathbf{n}^{\prime}_\mathrm{R}$ is the white additive Gaussian noise before the passive PS splitter whose elements are i.i.d. following $\mathcal{CN}(0,\sigma^{\prime 2}_{\mathrm{R}})$. 

The relay harvests part of the energy of the received signal through its PS receiver. We denote the PS factor by $\rho \in (0,1)$. The received signal power at the relay  is split into a ratio of $\rho
:1-\rho$ for ID and EH, receptively. Hence, the information for decoding at the relay  is as follows,
\begin{align}
\mathbf{y}_{\mathrm{ID}}&=\sqrt{\rho}(\sqrt{p_s}\mathbf{G}_S\mathbf{x}+\mathbf{n}^{\prime}_\mathrm{R})+\mathbf{n}^{\prime\prime}_{\mathrm{R}},\notag
\end{align}
where $\mathbf{n}^{\prime\prime}_{\mathrm{R}}$ is the white additive  ID noise after the passive PS with $\mathcal{CN}(0,\sigma^{\prime\prime 2}_{\mathrm{R}})$ elements. We define $\mathbf{n}_{\mathrm{R}}=\sqrt{\rho}\mathbf{n}^{\prime}_\mathrm{R}+\mathbf{n}^{\prime\prime}_{\mathrm{R}}$, which is the  effective additive noise vector at the relay. It can be seen that elements of $\mathbf{n}_{\mathrm{R}}$ follow  $\mathcal{CN}(0,\sigma^2_{\mathrm{R}})$, where $\sigma^2_{\mathrm{R}}=\rho\sigma^{\prime 2}_{\mathrm{R}}+\sigma^{\prime\prime 2}_{\mathrm{R}}$. Therefore, 
\begin{align}
\mathbf{y}_{\mathrm{ID}}&=\sqrt{\rho p_s}\mathbf{G}_S\mathbf{x}+\mathbf{n}_{\mathrm{R}},\notag\\
&=\sqrt{\rho p_s}\mathbf{g}_{Sk}{x_k}+\sqrt{\rho p_s}\sum_{j\ne k}^{K}\mathbf{g}_{Sj}x_j+\mathbf{n}_{\mathrm{R}}.\notag
\end{align} 
{\color{black}Notice that for a matrix $\mathbf{A}$, $\mathbf{a}_{i}$ denotes the $i$th column of $\mathbf{A}$. 
}

The portion  of the received signal for EH is
\begin{align}
	\sqrt{1-\rho}(\sqrt{p_s}\mathbf{G}_S\mathbf{x}+\mathbf{n}^{\prime}_\mathrm{R}),
	\notag
\end{align}
$\mathbf{n}^{\prime}_\mathrm{R}$ cannot be harvested and  the noise power at the relay EH receiver is ignored. Thus, the received signal for EH at the relay is approximated as
\begin{align}
\sqrt{(1-\rho){p_s}}\mathbf{G}_S\mathbf{x}. \label{eq:EH_app}
\end{align}
 A closed-form expression for the harvested energy at the relay is found in the following lemma.

 \begin{Lemma}\label{lem:P_R}
The harvested energy at the relay per unit time is
	\begin{align}
	E_\mathrm{H}&=\eta(1-\rho) p_sN\sum_{i=1}^{K}\beta_{Si}(\theta_{\mathrm{tilt}}),\label{eq:EH_final}
	\end{align}
	where $\eta$ is the RF-to-DC conversion efficiency.

\textit{Proof:} Please see Appendix \ref{App:proof_P_R}.
\end{Lemma}
 When the relay receives the source users' information, it performs MRC to maximize the received signal power, then, adopts the DF scheme. The MRC beam-forming matrix is $\mathbf{A}=\mathbf{G}_S^H$. The signal vector at the relay after MRC is denoted by $\mathbf{r}^{\mathrm{R}}$, where the $k$th entry pertains to the $k$th user information as the following,
\begin{align}
{r}_{k}^{\mathrm{R}}&=\sqrt{\rho p_s}\|\mathbf{g}_{Sk}\|^2{x_k}+\sqrt{\rho p_s}\mathbf{g}_{Sk}^H\sum_{j\ne k}^{K}\mathbf{g}_{Sj}x_j+\mathbf{g}_{Sk}^H\mathbf{n}_{\mathrm{R}}.\label{eq:kth_user}
\end{align}
For the achievable sum-rate analysis, we use the method in \cite{hassibi2003much} 
for mMIMO systems to find a lower bound, which is shown to be tight when $N$ approaches infinity. The key is to  
write the channel vector norm  in the received signal term as the summation of a known mean and the difference part. Hence, 
\begin{align}
{r}_{k}^{\mathrm{R}}&=\sqrt{\rho p_s}\mathbb{E}\{\|\mathbf{g}_{Sk}\|^2\}{x_k}+
\sqrt{\rho p_s}(\|\mathbf{g}_{Sk}\|^2-\mathbb{E}\{\|\mathbf{g}_{Sk}\|^2\}){x_k}\notag\\
&+\sqrt{\rho p_s}\sum_{j\ne k}^{K}\mathbf{g}_{Sk}^H\mathbf{g}_{Sj}x_j+\mathbf{g}_{Sk}^H\mathbf{n}_{\mathrm{R}}.\label{eq:det_r_k}
\end{align}
By treating the second term as noise, the average SINR of  $u_{Sk}$ at the relay in Phase I, denoted by $\gamma_k^{\mathrm{MRC}}$, has the following lower bound:
\begin{align}
\gamma_{k}^{\mathrm{MAC}} \ge\bar{\gamma}^{\mathrm{MAC}}_{k}=\frac{  N\beta_{Sk}(\theta_{\rm{tilt}})}{ \beta_{Sk}(\theta_{\rm{tilt}})+ \sum_{j\ne k}^{K}\beta_{Sj}(\theta_{\rm{tilt}})+\frac{\sigma^2_\mathrm{R}}{\rho p_s}}.\label{eq:lemm_2}
\end{align}
 The calculations of $\bar{\gamma}_k^{\mathrm{MAC}}$ are provided in Appendix \ref{App:proof_rate_phaseI}. By following the analysis in \cite{hassibi2003much}, it can also be shown that $\gamma_{k}^{\mathrm{MAC}}\xrightarrow[N\rightarrow\infty]{\text{a.s.}} \bar{\gamma}^{\mathrm{MAC}}_{k}$, where $\xrightarrow[N\rightarrow\infty]{\text{a.s.}}$ means almost sure convergence when $N\rightarrow \infty$. Further, when $N \to\infty$, the normalized effective noise term  in (\ref{eq:det_r_k}), the sum of the $2$nd, $3$rd and $4$th terms, converges in distribution to Gaussian. Thus, for the average achievable rate of $u_{Sk}$ at the relay, we have 
\begin{align}
R^{\mathrm{MAC}}_{k}=\log_2(1+\mathrm{\gamma_{k}^\mathrm{MAC}})\xrightarrow[N\rightarrow\infty]{\text{a.s.}}\log_2(1+ \bar{\gamma}^{\mathrm{MAC}}_{k}). 
\end{align}   
\subsection{Phase II: Broadcast (BC)}
The second phase is the broadcast phase. In this phase,  
the relay uses  MRT  on the decoded information signals to design its transmit signal vector as the following 
\begin{align}
\mathbf{x}_{\mathrm{R}}=\sqrt{P_\mathrm{R}} \alpha\mathbf{G}_{D}^* \mathbf{x},
\end{align} 
where $\alpha$ is the power coefficient and $P_R={E_\mathrm{H}}$ is the relay transmission power as in (\ref{eq:EH_final}).
From the power constraint  at the relay, $\mathbb{E}\{|\mathbf{x}_{\mathrm{R}}|^2\}=P_\mathrm{R}$, we have
 \begin{align}
 \alpha={\frac{1}{\sqrt{\mathbb{E}\left\{\mathrm{tr}\left\{\mathbf{G}^T_{D}\mathbf{G}^*_D\right\}\right\}}}}={\frac{1}{\sqrt{N\sum_{i=1}^{K}\beta_{Di}(\theta_\mathrm{tilt})}}}.
 \end{align}
  By noticing that the channel matrix from the relay to the destination users is $\mathbf{G}_D^T$, the received signal at the $k$th destination user is
\begin{align}
{r}_{k}^{\mathrm{D}}&=\sqrt{P_\mathrm{R}}\alpha\|\mathbf{g}_{Dk}\|^2{x_k}+\sqrt{P_\mathrm{R}}\alpha\mathbf{g}_{Dk}^*\sum_{j\ne k}^{K}\mathbf{g}^T_{Dj}x_j+n^{\mathrm{D}}_k,\label{eq:dest_u_k}
\end{align}
where $n^{\mathrm{D}}_k$ is the white additive noise at the $k$th destination user with $\mathcal{CN}(0,\sigma^2_{\mathrm{D}})$ distribution. 
By following the same steps as in (\ref{eq:det_r_k}) and Appendix~\ref{App:proof_rate_phaseI}, the lower-bound expression for the average SINR at $u_{Dk}$ in Phase II is as shown in (\ref{eq:lemm_3}), where $\gamma_{k}^{\mathrm{BC}}\xrightarrow[N\rightarrow\infty]{\text{a.s.}} \bar{\gamma}^{\mathrm{BC}}_{k}$.
Then, for the average achievable rate from the relay to $u_{Dk}$, we have
\begin{align}
R^{\mathrm{BC}}_{k}=\log_2(1+\mathrm{\gamma_{k}^\mathrm{BC}})\xrightarrow[N\rightarrow\infty]{\text{a.s.}}\log_2(1+ \bar{\gamma}^{\mathrm{BC}}_{k}). 
\end{align}   

The average achievable rate of the $k$th pair of source and destination is given as
\begin{align}
R_k=\min\{R^{\mathrm{MAC}}_{k},R^{\mathrm{BC}}_{k}\}.
\end{align}
Finally, the average achievable sum-rate of all user pairs is 
\begin{align}
R_{\mathrm{sum}}=\sum_{i=1}^{K} R_{k}.
\end{align} \begin{figure*}[t]
	\begin{align}
\gamma_{k}^{\mathrm{BC}} \ge	\bar{\gamma}^{\mathrm{BC}}_{k}=\frac{  N\beta_{Dk}(\theta_{\rm{tilt}})}{ \beta_{Dk}(\theta_{\rm{tilt}})+ \sum_{j\ne k}^{K}\beta_{Dj}(\theta_{\rm{tilt}})+\frac{\sigma^2_\mathrm{D}\sum_{i=1}^{K}\beta_{Di}(\theta_{\mathrm{tilt}})}{\eta(1-\rho) p_sN\beta_{Dk}(\theta_{\rm{tilt}})\sum_{i=1}^{K}\beta_{Si}(\theta_{\mathrm{tilt}})}}\label{eq:lemm_3}
	\end{align}
	\vspace{-1em}
\end{figure*}
\subsection{Antenna Tilt and PS Ratio Optimization}
In order to improve SWIPT efficiency, the joint optimization of the tilt and PS ratio is proposed aiming at maximizing the average achievable sum-rate of all user pairs which is formulated as below,
\vspace{-1em}
\begin{align}
&\max_{\theta_{\mathrm{tilt}},\rho}\hspace{1cm} R_{\mathrm{sum}}\\
&\hspace{1mm}\text{s.t.}\hspace{1cm} 0<\rho<1 \notag \\
& \hspace{1cm} 0<\theta_{\rm{tilt}}<\pi/2 \notag
\end{align}
For this non-convex problem, we use the grid search algorithm, due to the following reasons. First,
 the complex relation between $R_{\mathrm{sum}}$, $\theta_{\mathrm{tilt}}$, and $\rho$ makes finding a closed-form expression for the optimum $\rho$ and $\theta_{\mathrm{tilt}}$ not possible.
Second, due to the small ranges of values that $\rho$ and $\theta_{\mathrm{tilt}}$ can take, the grid search algorithm works well.

 \label{Sec:Sys_model}

	\section{Simulation Results}
\label{Sec:Simu}
In this section, we show simulation results on the average achievable sum-rate. The parameters of the relay antenna array gain in (\ref{eq:antenna_gain}) are defined previously in Section \ref{sec:sys_antenna}, and more parameter configurations are listed in Table \ref{Tab:parameters}.

\begin{figure}[t]	\centering
 	\includegraphics[width=8.5cm, height=6.38cm]{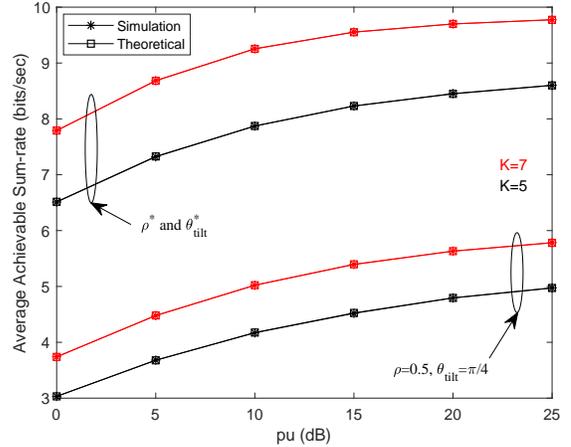}
 	\vspace{-1.2em}
 	\caption{Average sum-rate versus average user power for two designs: 1) $\rho=0.5$, $\theta_{\mathrm{tilt}}=\pi/4$, and   2) the optimal design $\rho^*$, $\theta_{\mathrm{tilt}}^*$.} 
 	\label{fig:pu_K}
 	\vspace{-1em}
 \end{figure}

\begin{figure}[t]
 	\centering
 	\includegraphics[width=8.5cm, height=6.38cm]{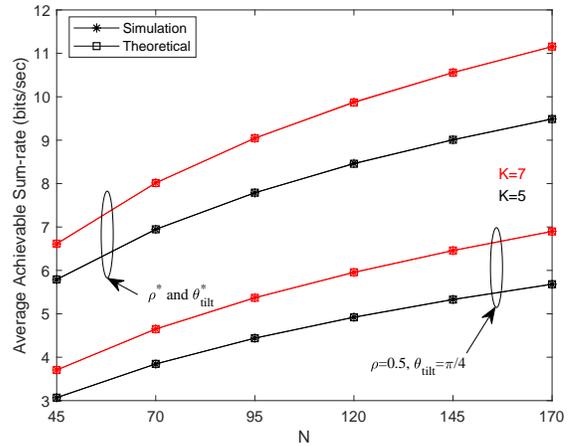}
 	\vspace{-1.2em}
 	\caption{Average sum-rate versus the number of relay antennas for two designs: 1) $\rho=0.5$, $\theta_{\mathrm{tilt}}=\pi/4$, and 2) the optimal design $\rho^*$, $\theta_{\mathrm{tilt}}^*$.}
 	\label{fig:N_k}
 	\vspace{-1em}
 \end{figure}

\begin{table}[H]
	\vspace{-1.5em} 
	\centering
	\caption{Simulation parameter values}
	\begin{tabular}{| c|c |} 
		\hline
		Parameters & Values \\
		\hline
		Number of pairs of users, $K$& 5, 7 \\
		Energy conversion efficiency at EH, $\eta$ & 0.5 \\ 
		Noise power at ID, $\sigma^2_{\mathrm{R}}$ & $-70$ dB \\ 
		Noise power at destination users, $\sigma^2_{\mathrm{D}}$ &  $-50$ dB \\ 
		Path loss exponent, $\nu$ & $3.76$\\
		Maximum distance between relay and users & $10$ m\\
		Search step size for $\theta_{\mathrm{tilt}}$ & $0.1047$ rad\\
		Search step size for $\rho$ & $0.0667$\\
		\hline
	\end{tabular}
	\label{Tab:parameters}
	\vspace{-1em}
\end{table}
 In Fig. \ref{fig:pu_K} and \ref{fig:N_k}, we show the theoretical and simulation results for the average sum-rate versus the average user power, and the number of antennas, respectively. Two designs are simulated: 1) $\rho=0.5$, $\theta_{\mathrm{tilt}}=\pi/4$, which are the middle points of the possible ranges for $\rho$ and $\theta_{\mathrm{tilt}}$, respectively, and 2) the optimum PS ratio, $\rho^*$, and tilt angle,  $\theta^*_{\mathrm{tilt}}$. Fig. \ref{fig:pu_K} shows the sum-rate results when $p_u$ changes form $0$ to $25$ dB, $N=100$, and $K=5,7$. Fig. \ref{fig:N_k} presents the sum-rate results when $N$ changes from $45$ to $170$, $p_u=15$ dB, and $K=5,7$. Each point of the figures is obtained by averaging over the sum-rates for $100$ randomly generated location sets for the $K$ user pairs. For the optimal design, the optimization is performed over the theoretical sum-rate for each location set. The results are then verified by the Monte-Carlo simulations where for each random location set, $10^3$ channel realizations are generated.
  The two figures show that the theoretical results perfectly match the Monte-Carlo simulations for all user power and relay antenna number ranges.

 Fig. \ref{fig:pu_K} also shows that the sum-rate is an increasing function of $p_u$ with a negative acceleration. Further, it reports that compared to the ordinary case  of $\rho=0.5$, $\theta_{\mathrm{tilt}}=\pi/4$, the proposed optimum system can bring the average sum-rate improvements of at least $72.95\%$ for $K=5$, and $68.97\%$ for $K=7$, both of which happens when $p_u=25$ dB. It also indicates that by increasing $p_u$ the sum-rate improvement due to optimization increases. Finally, this figure indicates that a self-sufficient energy harvesting relay with $100$ antennas can provide an average sum-rate of $6.5$ bits/sec for $5$ pairs of users with only $1$ watt user power.

 Fig. \ref{fig:N_k} shows that the sum-rate is an increasing function of $N$ with an almost linear relation. It reports that compared to the ordinary case  of $\rho=0.5$, $\theta_{\mathrm{tilt}}=\pi/4$, the proposed optimum system can bring the average sum-rate improvements of at least $67.04\%$  for $K=5$, and $61.81\%$ for $K=7$, both of which happens when $N=170$. It also indicates that by increasing $N$ the sum-rate improvement due to optimization increases. 
 \vspace{-1em}


	\vspace{-0.5em}
\section{Conclusion}
\label{Sec:Conc} 
In this paper, we examine a 3D mMIMO multi-user one-way relay network where the relay harvests energy from the users based on SWIPT PS scheme. We have derived an exact closed-form lower-bound expression for the average SINR using results for Haar matrices. Based on that, a tight closed-form approximation for the ergodic achievable rate of each pair of users is derived.
Further, a joint optimization problem over the PS ratio and the relay tilt is formulated to maximize the average achievable sum-rate. Our simulation results show the significant performance improvement provided by the extra degree-of-freedom offered by the vertical tilt and SWIPT.

	\appendices

\section{Proof of Lemma \ref{lem:P_R}} \label{App:proof_P_R}
From (\ref{eq:EH_app}), the harvested energy at the relay per unit time is,
\begin{align}
E_\mathrm{H}&=\eta(1-\rho) p_{s}\mathbb{E}\{\mathrm{tr}\{\mathbf{G}_{S}\mathbf{G}_{S}^H\}\}.\label{eq:EH_1}
\end{align} 
We consider the following singular-value decomposition (SVD):
\begin{align*}
\mathbf{H}_S=\mathbf{U}\mathbf{\Sigma}\mathbf{V}^H, 
\end{align*}
where $\mathbf{U}$, $\mathbf{V}$, and $\mathbf{\Sigma}$  are $N\times K$, $K \times K$, and $K\times K$ matrices.  $\mathbf{U}$ and $\mathbf{V}$  contain the  singular vectors of $\mathbf{H}_S$ and $\mathbf{\Sigma}=\mathrm{diag}\{\sigma_1, \sigma_2,\cdots, \sigma_K\}$  contains the singular values of $\mathbf{H}_S$. According to
Definition 2.5 in \cite{tulino2004random}, $\mathbf{U}$ and $\mathbf{V}$ are Haar matrices. Also, since entries of $\mathbf{H}_S$  follow i.i.d. $\mathcal{CN}(0, 1),$ $\mathbf{U}$, $\mathbf{V}$, and $\mathbf{\Sigma}$  are independent.
Therefore,
\begin{align}
\mathbb{E}\{\mathrm{tr}\{\mathbf{G}_{S}\mathbf{G}_{S}^H\}\}&=\mathbb{E}\{\mathrm{tr}\{\mathbf{H}_S\mathbf{D}_S\mathbf{H}^H_S\}\}\notag\\
&=\mathbb{E}\{\mathrm{tr}\{\mathbf{\Sigma}^2\mathbf{V}^H\mathbf{D}_S\mathbf{V}\}\}\notag\\
&=\sum_{i_2=1}^{K}\mathbb{E}\{\sigma_{i_2}^2\}\sum_{i_1=1}^{K}\beta_{Si_1}(\theta_{\mathrm{tilt}})\mathbb{E}\{|v_{i_1i_2}|^2\}.\notag
\end{align}  
{\color{black}Notice that for a matrix $\mathbf{A}$, $a_{ij}$ denotes the $(i,j)$th entry of $\mathbf{A}$.} 
According to \cite{hiai1999asymptotic}, for  any $i,j$, 
\begin{align}\mathbb{E}\{|v_{ij}|^2\}=\frac{1}{K}.\notag
\end{align}Further, for any arbitrary $i$,  we have 
\begin{align}
\mathbb{E}\{\sigma_i^2\}=\frac{1}{K}\mathbb{E}\{\mathrm{tr}\{\mathbf{H}_S^H\mathbf{H}_S\}\}=N.
\end{align}
 Therefore, (\ref{eq:EH_final}) is resulted.
\section{Proof of Equation (\ref{eq:lemm_2})} \label{App:proof_rate_phaseI}
According to (\ref{eq:det_r_k}),
 \begin{align}
 \bar{\gamma}^{\mathrm{MAC}}_{k}\hspace{-2mm}=\frac{|\mathbb{E}\{|\mathbf{g}_{Sk}|^2\}|^2}{\Var\{|\mathbf{g}_{Sk}|^2\}+\sum_{j\ne k}^{K}\mathbb{E}\{\mathbf{g}_{Sk}^H\mathbf{g}_{Sj}\mathbf{g}_{Sj}^H\mathbf{g}_{Sk}\}+\frac{\mathbb{E}\{|\mathbf{g}_{Sk}|^2\}\sigma^2_{\mathrm{R}}}{\rho p_s}}.\notag
 \end{align}
It can be shown that
\begin{align}
\mathbb{E}\{|\mathbf{g}_{Sk}|^2\}&= N\beta_{Sk}(\theta_{\rm{tilt}}), \label{eq:num_1}\\
\Var\{|\mathbf{g}_{Sk}|^2\}&=\mathbb{E}\{|\mathbf{g}_{Sk}|^4\}-|\mathbb{E}\{|\mathbf{g}_{Sk}|^2\}|^2,\notag
\end{align}
and
\begin{align}
\mathbb{E}\{|\mathbf{g}_{Sk}|^4\}&=\beta^2_{Sk}(\theta_{\rm{tilt}})\mathbb{E}\left\{|\mathbf{h}_{Sk}|^4\right\}\notag\\
&=\beta^2_{Sk}(\theta_{\rm{tilt}})\mathbb{E}\left\{\left(\sum_{i=1}^{N}|h_{Sik}|^2\right)^{\hspace{-2mm}2}\right\}\notag\\
&=\beta^2_{Sk}(\theta_{\rm{tilt}})\mathbb{E}\left\{\hspace{-1mm}\sum_{i=1}^{N}|{h}_{Sik}|^4\hspace{-1mm}+\hspace{-1mm}\sum_{i\ne j}^{N}\sum_{j=1}^{N}|{h}_{Sik}|^2|{h}_{Sjk}|^2\hspace{-1mm}\right\}.\notag
\end{align}
For any $i,k$, we can write  ${h}_{Sik}={h}_\mathrm{R}+j{h}_{\mathrm{I}}$,  where $ {h}_\mathrm{R}$ and $ {h}_{\mathrm{I}}$ are the real and imaginary components that are i.i.d. following $\mathcal{N}(0,1/2)$. Also,
\begin{align}
 \mathbb{E}\{{h}^4_\mathrm{R}\}=\mathbb{E}\{{h}^4_\mathrm{I}\}=3/4. \notag
\end{align}Thus,
\begin{align}
\mathbb{E}\left\{|{h}_{Sik}|^4 \right\} &=\mathbb{E}\{{h}^4_\mathrm{R}+{h}^4_{\mathrm{I}}+2{h}^2_\mathrm{R}{h}^2_{\mathrm{I}}\}=2,\notag\\
\Var\{|\mathbf{g}_{Sk}|^2\}&=N\beta^2_{Sk}(\theta_{\rm{tilt}}).  \label{eq:den_1}
\end{align} 
Further,  for any $k,j$, 
\begin{align}
\sum_{j\ne k}^{K}\hspace{-0.5mm}\mathbb{E}\hspace{-1mm}\left\{\mathbf{g}_{Sk}^H\mathbf{g}_{Sj}\mathbf{g}_{Sj}^H\mathbf{g}_{Sk}\right\}\hspace{-1mm}&=\hspace{-1mm}\beta_{Sk}(\theta_{\mathrm{tilt}})\hspace{-1.5mm}\sum_{j\ne k}^{K}\hspace{-0.5mm}\beta_{Sj}(\theta_{\mathrm{tilt}})\hspace{-0.3mm}\mathbb{E}\hspace{-0.3mm}\left\{\mathbf{h}_{Sk}^H\mathbf{h}_{Sj}\mathbf{h}_{Sj}^H\mathbf{h}_{Sk}\right\}\hspace{-1mm}\notag \label{eq:den_2}
\end{align}
and 
\begin{align}
\mathbb{E}\left\{\mathbf{h}_{Sk}^H\mathbf{h}_{Sj}\mathbf{h}_{Sj}^H\mathbf{h}_{Sk}\right\}=\mathbb{E}
\left\{\mathbf{h}_{Sk}^H\mathbf{h}_{Sk}\right\}=N.
\end{align}
Hence, by using (\ref{eq:num_1}), (\ref{eq:den_1}), and (\ref{eq:den_2}) in $\bar{\gamma}^{\mathrm{MAC}}_{k}$, (\ref{eq:lemm_2}) results.

	\bibliographystyle{IEEEtran}
        \bibliography{IEEEexample}
\end{document}